\documentclass[10pt,english,floatfix,nofootinbib,superscriptaddress,aps,prd,preprint,twocolumn]{revtex4}
\usepackage[utf8]{inputenc}
\usepackage{float}
\usepackage{array}
\usepackage{lipsum}
\usepackage{bbm}
\usepackage{dsfont}
\usepackage{graphicx}
\usepackage{amsmath}
\usepackage{tikz}
\usepackage{makecell}
\usepackage{multirow}
\usepackage{braket}
\usepackage{bbm}
\usepackage{xcolor}
\usepackage{xcolor}
\usepackage[normalem]{ulem}

\DeclareRobustCommand{\redstrike}{%
  \bgroup
  \markoverwith{{\rule[0.55ex]{2pt}{0.4pt}}}%
  \ULon
}

\usetikzlibrary{quotes,angles}
\usetikzlibrary{arrows} 
\usetikzlibrary{decorations.markings}
\usepackage{graphicx}
\usepackage[english]{babel}
\usepackage{color}
\usepackage{subfig}
\usepackage{caption}
\usepackage{tensor}
\usepackage{esint}
\usepackage[dvips]{epsfig}
\usepackage[dvips]{graphicx}
\usepackage{float}
\usepackage{units}
\usepackage{textcomp}
\usepackage{mathrsfs}
\usepackage{amsmath}
\usepackage[makeroom]{cancel}
\usepackage{amssymb}
\usepackage{amsbsy}
\usepackage{amsfonts}
\usepackage{amssymb,mathrsfs,xcolor}
\usepackage{esint}
\usepackage{array}
\usepackage{graphicx}

\usepackage[a4paper, margin=1.4cm]{geometry}

\usepackage{hyperref}
\hypersetup{
    colorlinks,
    citecolor=blue,
    filecolor=green,
    linkcolor=purple,
    urlcolor=red,
}

\usepackage{slashed}

\newcommand{\ie}{\begin{equation}}
\newcommand{\fe}{\end{equation}}
\newcommand{\se}{\begin{eqnarray}}
\newcommand{\ff}{\end{eqnarray}}

\usepackage{hyperref}
\hypersetup{colorlinks,breaklinks,
			citecolor=[rgb]{0,0.0,1.0},
            urlcolor=[rgb]{0.0,0.0,1.0},
            linkcolor=[rgb]{0,0.5,0.9}}

\begin{document}

\title{Neutrino dynamics in a non--commutative spacetime}

\author{A. A. Ara\'{u}jo Filho}
\email{dilto@fisica.ufc.br}
\affiliation{Departamento de Física, Universidade Federal da Paraíba, Caixa Postal 5008, 58051--970, João Pessoa, Paraíba,  Brazil.}
\affiliation{Departamento de Física, Universidade Federal de Campina Grande Caixa Postal 10071, 58429-900 Campina Grande, Paraíba, Brazil.}
\affiliation{Center for Theoretical Physics, Khazar University, 41 Mehseti Street, Baku, AZ-1096, Azerbaijan.}

\author{N. Heidari}
\email{heidari.n@gmail.com}
\affiliation{Center for Theoretical Physics, Khazar University, 41 Mehseti Street, Baku, AZ-1096, Azerbaijan.}
\affiliation{School of Physics, Damghan University, Damghan, 3671641167, Iran.}

\author{Yuxuan Shi}
\email{shiyx2280771974@gmail.com}
\affiliation{Department of Physics, East China University of Science and Technology, Shanghai 200237, China}

\author{Iarley P. Lobo}
\email{lobofisica@gmail.com}
\affiliation{Departamento de Física, Universidade Federal de Campina Grande Caixa Postal 10071, 58429-900 Campina Grande, Paraíba, Brazil.}
\affiliation{Department of Chemistry and Physics, Federal University of Para\'iba, Rodovia BR 079 - km 12, 58397-000 Areia-PB,  Brazil.}


\date{\today}

\begin{abstract}

We investigate neutrino dynamics in a spherically symmetric non--commutative black hole spacetime generated by a Lorentzian smeared matter distribution. We work with the leading--order Lorentzian--smeared metric, in which the non--commutative parameter $\Theta$ enters explicitly through a $\sqrt{\Theta}$ correction to the Schwarzschild geometry. Within this background, we analyze the neutrino--antineutrino annihilation process, the oscillation phase acquired by massive neutrinos along geodesics, and the flavor--transition probability in the presence of weak gravitational lensing. We show that non--commutativity monotonically suppresses the relativistic energy deposition rate relative to the Schwarzschild case, while the Newtonian behavior is recovered at large radii. In the oscillation sector, the weak--field phase receives explicit $\sqrt{\Theta}$ corrections, which modify the distance of closest approach and reduce the accumulated phase with respect to the Schwarzschild limit. For lensed trajectories, the macroscopic transition probabilities remain close to the Schwarzschild prediction; however, a high--precision residual analysis reveals a coherent, non--vanishing oscillatory correction driven by the non--commutative geometry. The dependence on the absolute neutrino mass scale enters only through multi path interference terms and remains subdominant compared with the $\Theta$--induced phase shift. The full three--flavor residual analysis confirms that the same non--commutative oscillatory pattern persists beyond the effective two--flavor approximation.

\end{abstract}


\maketitle


\section{Introduction}

Black hole models have been extensively revised under frameworks where spacetime coordinates no longer commute. In these formulations, position operators satisfy $[x^\mu, x^\nu] = \mathbbm{i} \Theta^{\mu \nu}$, introducing a fixed antisymmetric matrix $\Theta^{\mu \nu}$ that alters the geometric foundation of the theory. Such non--commutative deformations have produced consequences across multiple aspects of black hole physics. Corrections to the evaporation lifetime profile \cite{myung2007thermodynamics,23araujo2023thermodynamics}, shifts in thermodynamic relations including entropy and specific heat \cite{nozari2007thermodynamics,banerjee2008noncommutative,nozari2006reissner,sharif2011thermodynamics,Heidari:2025iiv,araujo20sss25non}, and changes to the spectrum of quasinormal oscillations \cite{zhao2023quasinormal,anacleto2023absorption,Anacleto:2019tdj,heidari2024exploring,asdasd1,mann2011cosmological,campos2022quasinormal,anacleto2021quasinormal,asdasd2,karimabadi2020non,lopez2006towards,modesto2010charged,nicolini2009noncommutative}.

To make these deformations compatible with gravity, symmetry structures such as the Poincaré and de Sitter groups are extended using tools like the Seiberg--Witten map, which preserves gauge invariance in non--commutative coordinates \cite{Herceg:2023pmc}. Within this framework, the first non--commutative modification of the Schwarzschild solution was developed by Chaichian et al. \cite{chaichian2008corrections}, and a more recent extension of this construction was introduced in \cite{Juric:2025kjl}.

Instead of altering the geometric side of Einstein’s equations, one can introduce non--commutative effects by modifying the matter sector alone. This line of thought was explored by Nicolini et al. \cite{nicolini2006noncommutative}, who modeled the gravitational source not as a point mass but as a smeared energy distribution. The mass is spread out over a minimal length scale determined by the non--commutative parameter $\Theta$, leading to two representative density profiles: a Gaussian form, $\rho_\Theta = M (4\pi \Theta)^{-3/2} e^{-r^2 / 4\Theta}$, and a Lorentzian alternative, $\rho_\Theta = M \sqrt{\Theta} \pi^{-3/2} (r^2 + \pi \Theta)^{-2}$.

Neutrinos, due to their unique behavior and elusive nature, have remained a subject of intense study within particle physics \cite{neu42,neu43,neu44}. Unlike many other particles, the states through which neutrinos interact—called flavor states—do not coincide with their mass eigenstates. This misalignment leads to quantum interference effects, known as neutrino oscillations, where a neutrino created in one flavor can be detected as another after traveling some distance \cite{neu39,neu40,neu41}.

In a flat spacetime framework, these oscillations are governed by the squared differences between the neutrino masses, such as $
|\Delta m^2_{21}|, \quad |\Delta m^2_{31}|, \quad \text{and} \quad |\Delta m^2_{23}|$,  
with the general form $
\Delta m^2_{ij} = m^2_i - m^2_j$.  
Crucially, the transition probabilities derived in this context depend only on these differences and provide no direct access to the absolute mass values \cite{neu45}.

However, when neutrinos propagate through curved spacetimes, the situation changes. Gravitational effects modify the oscillation dynamics in a way that can, in principle, make them sensitive to absolute mass values {once multiple lensed trajectories interfere}. In such settings, the curvature of spacetime alters phase evolution and can introduce new contributions to the oscillation formula. This gravitational sensitivity becomes particularly important in the analysis of high--energy neutrinos originating from distant astrophysical events. By comparing observed flavor compositions with theoretical predictions, one may extract information not only about neutrino properties but also about the gravitational fields they have encountered \cite{neu46,neu47,neu48,neu49,neu50,neu51,neu52,neu53,Shi:2023hbw}.

Neutrino flavor transitions can be significantly influenced by the geometry of the spacetime through which they propagate. When treated from a geometric standpoint, the evolution of neutrino phases along geodesic paths offers a direct probe of the surrounding gravitational field \cite{neu54,neu55,Shi:2025rfq,Shi:2025plr}. In regions where spacetime curvature is strong, such as near compact astrophysical objects, the deflection of neutrino trajectories due to gravitational lensing can cause them to intersect or focus, modifying interference patterns and consequently altering oscillation probabilities \cite{neu53,Shi:2024flw}.

The investigation of neutrino behavior near such focal regions has gained attention in recent years, with several works analyzing how lensing--induced convergence affects flavor transitions \cite{neu56,neu57,neu58}. In rotating spacetimes, additional complexity arises: Swami demonstrated that the angular momentum of the gravitational source changes the phase evolution of neutrinos, which can enhance or suppress oscillation probabilities depending on the configuration—an effect that becomes particularly relevant for solar--scale systems \cite{neu59}.

Moreover, deviations from spherical symmetry have also been explored. Studies involving axially symmetric spacetimes, governed by a deformation parameter $\gamma$, reveal that even in static and asymptotically flat backgrounds, the presence of such a parameter leads to modifications in oscillation behavior. In these cases, the deformation can introduce a dependence on absolute neutrino masses—an outcome not present in flat spacetime analyses \cite{neu60}.

In this paper, {we examine} the effects of non--commutativity on neutrino dynamics. We first consider the energy deposition rate from the neutrino annihilation process. Next, we analyze the neutrino oscillation phase and transition probability within this framework. Finally, we explore the influence of non--commutativity on neutrino gravitational lensing.

\section{The black hole solution}

Spacetime structure can be revisited through non--commutative extensions of general relativity, as explored in several gravitational scenarios 
\cite{Anacleto:2019tdj,anacleto2023absorption,k10,k101,k102,k103,k6,k7,k8,k9,campos2022quasinormal}. 
One frequent approach introduces non-locality via the Moyal product, leading to modified field theories \cite{k11}. 
{
Here, however, we follow the complementary matter--sector implementation of non--commutativity, in which the point--like source is replaced by an extended distribution controlled by the minimal length scale $\sqrt{\Theta}$. 
}
In this way, this section starts by analyzing the specific black hole configuration under consideration, focusing on the smeared matter distribution given below 
\cite{campos2022quasinormal,nicolini2006noncommutative,nozari2008hawking,aa2025particle,Filho:2024zxx,AraujoFilho:2024rss}
\ie
\rho^{(\Theta)}(r) = \frac{M \sqrt{\Theta}}{\pi^{3/2} (r^{2} + \pi \Theta)^{2}},
\fe
in which $M$ denotes the mass and $\Theta$ represents the non--commutative parameter with the dimension of $[\mathrm{L}^{2}]$, given through the commutation relation 
\ie
[x^{\mu}, x^{\nu}] = i\Theta^{\mu\nu}. 
\fe

{
For this Lorentzian profile, the corresponding mixed component of the stress--energy tensor is taken as $T^{0}{}_{0}=-\rho^{(\Theta)}(r)$. Therefore, instead of postulating the metric function directly, one may obtain it from the Einstein equations by introducing the mass function $m_{\Theta}(r)$. For a static and spherically symmetric geometry written in Schwarzschild--like form,
} 
{ \ie \mathrm{d}s^{2} = -F_{\Theta}(r)\mathrm{d}\tau^{2} +\frac{1}{F_{\Theta}(r)}\mathrm{d}r^{2} +r^{2}\mathrm{d}\theta^{2} +r^{2}\sin^{2}\theta\,\mathrm{d}\varphi^{2}, \fe } 
{ with } 
{ \ie F_{\Theta}(r)=1-\frac{2m_{\Theta}(r)}{r}, \fe } 
{ the $(0,0)$ component of the Einstein equations gives \cite{Wang:2024fiz} }
{ \ie G^{0}{}_{0}=-\frac{2m_{\Theta}'(r)}{r^{2}} =8\pi T^{0}{}_{0}, \qquad m_{\Theta}'(r)=4\pi r^{2}\rho^{(\Theta)}(r). \fe } { In this way, the mass function is obtained as } { 
\ie
\begin{split}
m_{\Theta}(r) = & 4\pi\int_{0}^{r}\bar r^{\,2}\rho^{(\Theta)}(\bar r)\,\mathrm{d}\bar r \\
& = \frac{2M}{\pi} \arctan\left(\frac{r}{\sqrt{\pi\Theta}}\right) - \frac{2M\sqrt{\Theta}\,r}{\sqrt{\pi}\left(r^{2}+\pi\Theta\right)}. 
\end{split}
\fe } 
{ This expression satisfies $m_{\Theta}(r\rightarrow\infty)=M$, so that the spacetime is asymptotically Schwarzschild. The exact metric function generated by the Lorentzian distribution is therefore } { \ie f_{\Theta}(r) = 1 - \frac{4M}{\pi r} \arctan\left(\frac{r}{\sqrt{\pi\Theta}}\right) + \frac{4M\sqrt{\Theta}}{\sqrt{\pi}\left(r^{2}+\pi\Theta\right)}. \label{exact_lorentzian_metric} \fe } { The remaining components of the anisotropic source can be chosen consistently with this Schwarzschild--like gauge as  $p_{r}^{(\Theta)}=-\rho^{(\Theta)}$ and  $p_{\perp}^{(\Theta)}=-\rho^{(\Theta)}-(r/2)\mathrm{d}\rho^{(\Theta)}/\mathrm{d}r$, which ensures the conservation condition $\nabla_{\mu}T^{\mu}{}_{\nu}=0$. }

{
The metric function used in the present work is not the full expression in Eq. \eqref{exact_lorentzian_metric}. Instead, we adopt its large-distance expansion, valid for $r\gg\sqrt{\Theta}$, namely
}
{
\ie
f_{\Theta}(r)
=
1-\frac{2M}{r}
+\frac{8M\sqrt{\Theta}}{\sqrt{\pi}r^{2}}
-\frac{16M\sqrt{\pi}\,\Theta^{3/2}}{3r^{4}}
+\mathcal{O}\left(\frac{\Theta^{5/2}}{r^{6}}\right).
\fe
}
{
Keeping only the leading non--commutative correction, the effective geometry considered throughout the following sections is therefore
}
\ie
\label{noncomutativemetricmain}
\mathrm{d}s^{2} = -\mathrm{A}_{\Theta} (r) \mathrm{d}\tau^{2} 
+ \frac{1}{\mathrm{B}_{\Theta}(r)} \mathrm{d}r^{2} 
+ r^{2}\mathrm{d}\theta^{2} 
+ r^{2}\sin^{2}\theta \mathrm{d}\varphi^{2} ,
\fe
where
\ie
\label{metrreeedd}
{
\mathrm{A}_{\Theta}(r)=\mathrm{B}_{\Theta}(r)
\equiv f_{\Theta}(r)
=
1- \frac{2M}{r} + \frac{8M\sqrt{\Theta}}{\sqrt{\pi}r^{2}}
+\mathcal{O}\left(\Theta^{3/2}\right).
}
\fe
{
Thus, Eq. \eqref{metrreeedd} should be understood as the leading--order Lorentzian--smeared black hole metric. In the limit $\Theta\rightarrow0$, the mass distribution tends to a Dirac delta source and the Schwarzschild geometry is recovered.
}

{
At this order, the metric function has the same algebraic structure as the Reissner--Nordstr\"om solution and yields two physical horizons, provided that the discriminant is non-negative.}
They are 
\ie
r_{+} = M + \frac{\sqrt{\pi  M^2-8 \sqrt{\pi } \sqrt{\Theta } M}}{\sqrt{\pi }},
\fe
accounting for the event horizon and
\ie
r_{-} = M-\frac{\sqrt{\pi  M^2-8 \sqrt{\pi } \sqrt{\Theta } M}}{\sqrt{\pi }},
\fe
for the Cauchy horizon. 
{
These expressions are the roots of the truncated metric function \eqref{metrreeedd}; the exact horizon positions associated with Eq. \eqref{exact_lorentzian_metric} receive additional corrections from the omitted terms of order $\Theta^{3/2}$ and higher.
}

In the literature, there exists an alternative strategy {that} consists in absorbing the geometric corrections into an effective mass parameter, so that the modified solution retains a Schwarzschild--like form with a redefined mass. This mass--deformation prescription has been employed in non--commutative gauge theory and extended to both static 
\cite{heidari2024gravitational,AraujoFilho:2024mvz,heidari2024quantum} 
and axisymmetric configurations \cite{heidari2025axisymmetric}. 
{The perturbed metric has also been experimentally constrained using Solar System tests in \cite{Wang:2024fiz}, with best bounds at the centimeter scale $\sqrt{\Theta}\lessapprox 26\, \rm cm$ from the perihelion advance of Mercury.}
{
In contrast, the present analysis does not rely on a mass redefinition. We work with the leading-order Lorentzian-smeared metric \eqref{metrreeedd}, keeping the explicit $\Theta$--dependent correction while consistently neglecting the higher--order terms displayed in the expansion above.
}


\section{The energy deposition rate by the neutrino annihilation process}

The investigation focuses on how spacetime modified by mass deformation in a non--commutative black hole environment facilitates energy transfer. In this context, the annihilation of neutrino pairs serves as the primary channel for energy release. The rate at which this energy is deposited—quantified per unit volume and per unit time—is given by \cite{Salmonson:1999es}:
\ie
\dfrac{\mathrm{d}E(r)}{\mathrm{d}t\mathrm{d}V}=2KG_{f}^{2}f(r)\iint
n(\varepsilon_{\nu})n(\varepsilon_{\overline{\nu}})
(\varepsilon_{\nu} + \varepsilon_{\overline{\nu}})
\varepsilon_{\nu}^{3}\varepsilon_{\overline{\nu}}^{3}
\mathrm{d}\varepsilon_{\nu}\mathrm{d} \varepsilon_{\overline{\nu}}
\fe
where {$f(r)$ represents the term arising from the integration over angular variables, and}
\ie
K = \dfrac{1}{6\pi}(1\pm4\sin^{2}\theta_{W}+8\sin^{4} \theta_{W}).
\fe
Adopting the Weinberg angle value $\sin^{2}\theta_{W} = 0.23$, one obtains the corresponding expressions for different neutrino pair combinations as outlined in \cite{Salmonson:1999es}:
\ie
K(\nu_{\mu},\overline{\nu}_{\mu}) = K(\nu_{\tau},\overline{\nu}_{\tau})
=\dfrac{1}{6\pi}\left(1-4\sin^{2}\theta_{W} + 8\sin^{4}\theta_{W}\right)
\fe
and
\ie
K(\nu_{e},\overline{\nu}_{e})
=\dfrac{1}{6\pi}\left(1+4\sin^{2}\theta_{W} + 8\sin^{4}\theta_{W}\right)
\fe
respectively \cite{Salmonson:1999es}. The distinct expressions for each neutrino pair correspond to the choice $\sin^{2}\theta_{W} = 0.23$ for the Weinberg angle. The weak interaction strength is governed by the Fermi constant, taken to be $G_{f} = 5.29 \times 10^{-44} , \text{cm}^{2} , \text{MeV}^{-2}$. Under these assumptions, the term arising from the integration over angular variables is formulated in the following manner \cite{Salmonson:1999es}:
\begin{align}
f(r)&=\iint\left(1-\bm{\Omega_{\nu}}\cdot\bm{\Omega_{\overline{\nu}}}\right)^{2}
\mathrm{d}\Omega_{\nu}\mathrm{d}\Omega_{\overline{\nu}}\notag\\
&=\dfrac{2\pi^{2}}{3}(1 - x)^{4}\left(x^{2} + 4x + 5\right)
\end{align}
with
\ie
x = \sin\theta_{r}.
\fe

At a radial distance $r$, the angle $\theta_r$ measures how a particle’s trajectory deviates from the local tangential direction of a circular orbit. The directional motion of neutrinos and antineutrinos is described by the unit vectors $\Omega_{\nu}$ and $\Omega_{\overline{\nu}}$, with their respective differential solid angles denoted by $\mathrm{d}\Omega_{\nu}$ and $\mathrm{d}\Omega_{\overline{\nu}}$. When the system reaches thermal equilibrium at temperature $T$, the occupation numbers for neutrinos and antineutrinos in phase space, $n(\varepsilon_{\nu})$ and $n(\varepsilon_{\overline{\nu}})$, are determined by the Fermi--Dirac distribution function \cite{Salmonson:1999es}
\ie
n(\varepsilon_{\nu}) = \frac{2}{h^{3}}\dfrac{1}{e^{\left({\frac{\varepsilon_{\nu}}{k \, T}}\right)} + 1}.
\fe

In this context, $h$ stands for Planck’s constant and $k$ refers to Boltzmann’s constant. With these constants specified, the expression that quantifies the rate of energy deposition per unit volume and per unit time can be written as \cite{Salmonson:1999es}
\ie
\frac{\mathrm{d}E}{\mathrm{d}t\mathrm{d}V} = \frac{21\zeta(5)\pi^{4}}{h^{6}}K G_{f}^{2} f(r)(k \, T)^{9}.
\fe
The expression for $\mathrm{d}E/\mathrm{d}t \, \mathrm{d}V$ serves as a key element in the analysis of how energy is transformed within compact astrophysical objects \cite{Salmonson:1999es}. It incorporates the radial dependence of physical quantities, most notably the temperature profile $T = T(r)$, which defines the thermal state at each point in space \cite{Salmonson:1999es}.

As measured by an observer situated at radius $r$, the local temperature $T(r)$ obeys the redshift relation $T(r)\sqrt{\mathrm{A}_{\Theta}(r)} = \text{constant}$, reflecting the impact of the gravitational field on thermal measurements \cite{Salmonson:1999es}. At the surface of the neutrinosphere, the temperature associated with neutrino emission is specified by the following relation \cite{Salmonson:1999es}:
\ie
T(r)\sqrt{\mathrm{A}_{\Theta}(r)} = T(R)\sqrt{\mathrm{A}_{\Theta}(R)}.
\fe
Here, $R$ denotes the radial coordinate corresponding to the surface of the compact object acting as the gravitational source. To simplify subsequent computations, the local temperature $T(r)$ is replaced using the relation provided in identity (13). Taking into account gravitational redshift effects, the neutrino luminosity is expressed as \cite{Salmonson:1999es}:
\ie
L_{\infty} = \mathrm{A}_{\Theta}(R)L(R)
\fe
where the luminosity corresponding to a single neutrino flavor, being evaluated at the neutrinosphere, is given by \cite{Salmonson:1999es}:
\ie
L(R) = 4 \pi {R^{2}}\frac{7}{4}\frac{\Tilde{a}\,c}{4}T^{4}(R).
\fe

In this context, $\Tilde{a}$ denotes the radiation constant, while $c$ is the speed of light in vacuum. To rewrite the temperature as a function of the observer’s radial location, the following relation is used \cite{Salmonson:1999es}
\ie
\begin{split}
\frac{\mathrm{d}E(r)}{\mathrm{d}t \, \mathrm{d}V} & = \dfrac{21\zeta(5)\pi^{4}}{h^{6}}
KG_{f}^{2}k^{9}\left(\frac{7}{4}\pi \Tilde{a}\,c\right)^{-\frac{9}{4}}\\
& \times L_{\infty}^{\frac{9}{4}}f(r)
\frac{\left[\mathrm{A}_{\Theta}(R)\right]^{\frac{9}{4}}}{\left[\mathrm{A}_{\Theta}(r)\right]^{\frac{9}{4}}} R^{-\frac{9}{2}},
\end{split}
\fe
in which $\zeta(s)$ refers to the Riemann zeta function, mathematically represented for values $s > 1$ through the infinite summation: 
\ie
\zeta(s) = \sum_{n=1}^{\infty} \frac{1}{n^s},
\fe  
which converges for all real values of $s$ greater than $1$. It is important to mention that, beyond the dependence on the radial coordinate, the metric components evaluated at the surface of the compact source also influence the expression for the local energy deposition rate. To determine the total radiative energy output in the presence of gravity, one must integrate the energy deposition density over time. Calculating the angular contribution $f(r)$ requires a deeper investigation of the previously introduced variable $x$. This involves solving the null geodesic equations in the spacetime of a spherically symmetric mass distribution, as shown in \cite{Salmonson:1999es} and further discussed in \cite{Lambiase:2020iul,Shi:2023kid}
\begin{align}
x^{2}& = \sin^{2}\theta_{r}|_{\theta_{R}=0}\notag\\
&=1-\dfrac{R^{2}}{r^{2}}\dfrac{\mathrm{A}_{\Theta}(r)}{\mathrm{A}_{\Theta}(R)}.
\end{align}
The angular integration term is directly shaped by the structure of the spacetime metric. This connection allows one to evaluate the total energy deposited by integrating the local deposition rate—expressed per unit volume and per unit time—throughout the spherical region surrounding the central gravitational object \cite{Lambiase:2020iul,Shi:2023kid}
\ie
\begin{split}
\dot{Q} & = \frac{\mathrm{d}E}{\sqrt{\mathrm{A}_{\Theta}(r)}\mathrm{d}t}\\
&=\dfrac{84\zeta(5)\pi^{5}}{h^{6}}KG_{f}^{2}k^{9}
\left(\dfrac{7}{4}\pi \Tilde{a}\,c\right)^{-\frac{9}{4}}
L_{\infty}^{\frac{9}{4}}\left[\mathrm{A}_{\Theta}(R)\right]^{\frac{9}{4}}\\
& \times 
R^{-\frac{9}{2}}\int_{R}^{\infty}\frac{r^{2}f(r)}
{\mathrm{A}_{\Theta}(r)    \sqrt{-\mathrm{B}_{\Theta}(r) }   }\mathrm{d}r.
\end{split}
\fe

The symbol $\dot{Q}$ denotes the total rate at which neutrino energy is transformed into electron–positron pairs at a specific radial position \cite{Salmonson:1999es}. If this rate becomes sufficiently large, the resulting pair production can trigger explosive phenomena. To go further in this analysis, it is important to compare this relativistic energy deposition rate with its Newtonian counterpart \cite{Salmonson:1999es,Lambiase:2020iul,Shi:2023kid}.
\ie
\begin{split}
 \frac{\dot{Q}}{\dot{Q}_{\text{Newton}}} = 3\left[\mathrm{A}_{\Theta}(R)\right]^{\frac{9}{4}}
& \int_{1}^{\infty}(x - 1)^{4}\left(x^{2} + 4x + 5\right) \\
& \times \frac{y^{2}}
{\mathrm{A}_{\Theta}(Ry)^{\frac{9}{2}}  \sqrt{- \mathrm{B}_{\Theta}(Ry)} }\mathrm{d}y.
\end{split}
\fe

{By defining the dimensionless variable $\tilde{y}=r/R$ and substituting the metric functions $\mathrm{A}_{\Theta}(r)$ and $\mathrm{B}_{\Theta}(r)$ given in Eq.~(\ref{metrreeedd}), the radial derivative of the energy deposition rate, $\mathrm{d}\dot{Q}/\mathrm{d}r$, was recast explicitly as a function of $r$. This form allowed us to follow the spatial behavior of the energy deposition, highlighting its radial dependence and the emergence of possible amplification effects}
\ie
\begin{split}
\frac{\mathrm{d}\dot{Q}}{\mathrm{d}r} & = 4\pi\left(\frac{\mathrm{d}E}{\mathrm{d}t\,\mathrm{d}V}\right)\frac{1}{\sqrt{-\mathrm{B}_{\Theta}(r)}}r^{2}\\
&= \frac{168\zeta(5)\pi^{7}}{3h^{6}}KG_{f}^{2}k^{9}
\left(\dfrac{7}{4}\pi \Tilde{a}\,c\right)^{-\frac{9}{4}}
L_{\infty}^{\frac{9}{4}}\\
&\times(x-1)^{4}\left(x^{2} + 4x + 5\right)
\left[\frac{\mathrm{A}_{\Theta}(R)}{\mathrm{A}_{\Theta}(r)}\right]^{\frac{9}{4}} \\
& \times R^{-\frac{5}{2}}
\frac{1}{\sqrt{-\mathrm{B}_{\Theta}(r)}}\left(\frac{r}{R}\right)^{2}.
\end{split}
\fe

The quantity $\mathrm{d}\dot{Q}/\mathrm{d}r$ reflects the radial variation of the total energy deposition rate, measured outward from the center of the gravitational source, and explicitly incorporates the spacetime geometry through the metric functions. Examining how the internal structure of compact objects—especially within the framework of asymptotic safety—modifies neutrino-antineutrino annihilation is crucial for determining the scenarios in which such processes may trigger gamma--ray bursts. Following a sequence of algebraic steps, the resulting expression is given by \cite{Salmonson:1999es,Lambiase:2020iul,Shi:2023kid}:
\ie
\begin{split}
\dfrac{\dot{Q}}{\dot{Q}_{\text{Newton}}} = 3\left[\mathrm{A}_{\Theta}(R)\right]^{\frac{9}{4}}&\int_{1}^{\infty}
(x - 1)^{4}\left(x^{2} + 4x + 5\right)\\
& \times \frac{\Tilde{y}^{2}}{\left[\mathrm{A}_{\Theta}(Ry)\right]^{5}}\mathrm{d}\Tilde{y},
\end{split}
\fe
in which
{\ie
\begin{split}
\mathrm{A}_{\Theta}(R)&= f_{\Theta}(R) = 1 - \frac{2M}{R} + \frac{8M \sqrt{\Theta}}{\sqrt{\pi}R^{2}},\\
\mathrm{A}_{\Theta}(R\Tilde{y})&= f_{\Theta}(R\tilde{y})
= 1 - \frac{2M}{R}\,\frac{1}{\tilde{y}}
+ \frac{8M\sqrt{\Theta}}{\sqrt{\pi}\,R^{2}}\,\frac{1}{\tilde{y}^{2}}.
\end{split}
\fe
Also, in this manner, we can write
\ie
x^{2}= 1 - \frac{1}{\tilde{y}^{2}}
\frac{1 - \dfrac{2M}{R}\dfrac{1}{\tilde{y}} +
\dfrac{8M\sqrt{\Theta}}{\sqrt{\pi}\,R^{2}}\dfrac{1}{\tilde{y}^{2}}}{1 - \dfrac{2M}{R} +
\dfrac{8M\sqrt{\Theta}}{\sqrt{\pi}\,R^{2}}}\, .
\fe
}

\begin{figure}
\centering
\includegraphics[scale=0.55]{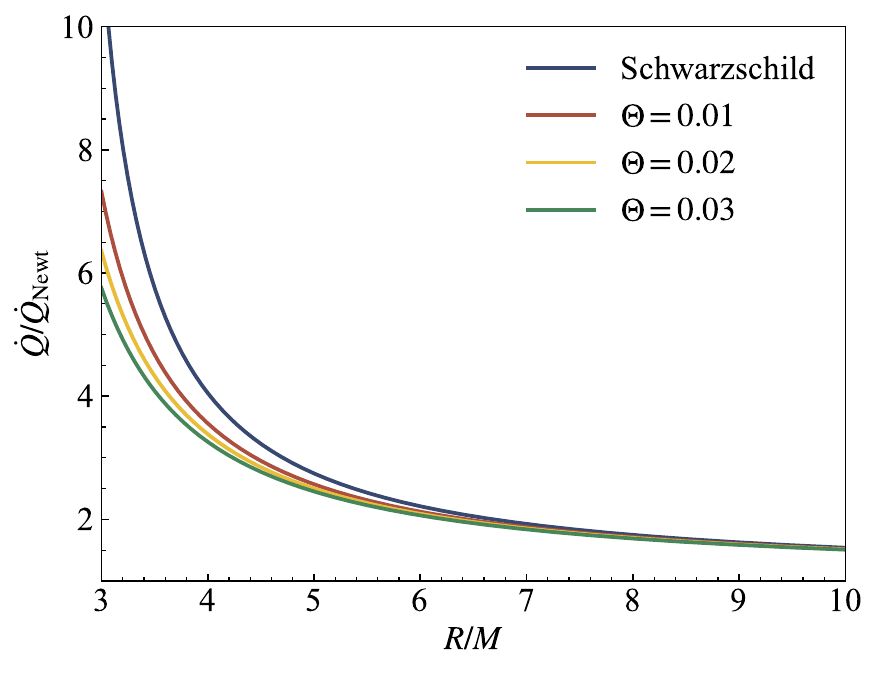}
\caption{{The quantity $\dot{Q}/\dot{Q}_{\text{Newton}}$ is shown as a function of $R/M$ for different values of $\Theta$.}}
\label{energydeposiiton}
\end{figure}

To facilitate the interpretation of the results, Fig.~\ref{energydeposiiton} presents a parametric plot of the ratio $\dot{Q}/\dot{Q}{\text{Newton}}$ as a function of ${R/M}$ for several values of the non--commutative parameter $\Theta$. The figure shows a monotonic decrease of $\dot{Q}/\dot{Q}{\text{Newton}}$ as $\Theta$ increases, indicating a systematic suppression of the energy deposition rate relative to its Newtonian counterpart. A quantitative assessment of this trend is reported in Tab.~\ref{tabbb}, where the corresponding numerical values are summarized. {Throughout this work, we adopt geometric units ($G=c=1$), in which all lengths are measured in units of the mass $M$, so that, since $\Theta$ has dimensions of $[\mathrm{L}^2]$, the numerical values of $\Theta$ quoted in the figures and tables represent the dimensionless ratio $\Theta/M^{2}$.}

An important aspect of the plot is the slow approach of the ratio toward its Newtonian limit, $\dot{Q}/\dot{Q}_{\text{Newton}}=1$. To make the asymptotic behavior explicit, the inset in Fig.~\ref{energydeposiiton} displays the ratio in the far--field interval $R/M\in[990,1000]$. In this regime, the curves corresponding to all values of $\Theta$ lie unambiguously close to unity, thereby demonstrating consistency with the expected Newtonian limit.


\begin{table}[h!]
\centering
\caption{\label{tabbb}The rate $\dot{Q}$ (in erg/s) for different values of $\Theta$ and $R/M$.}
\label{tab:dotQ}
\begin{tabular}{ccc}
\hline\hline
$\Theta$ & $R/M$ & $\dot{Q}$ \((\text{erg/s})\) \\
\hline
0.00 & 0 & $1.50 \times 10^{50}$ \\
\hline
\multirow{2}{*}{0.00} & 3 & $1.72 \times 10^{51}$ \\
                      & 4 & $0.61 \times 10^{51}$ \\
\hline
\multirow{2}{*}{0.01} & 3 & $1.10 \times 10^{51}$ \\
                      & 4 & $0.53 \times 10^{51}$ \\
\hline
\multirow{2}{*}{0.02} & 3 & $0.95 \times 10^{51}$ \\
                      & 4 & $0.51 \times 10^{51}$ \\
\hline
\multirow{2}{*}{0.03} & 3 & $0.86 \times 10^{51}$ \\
                      & 4 & $0.49 \times 10^{51}$ \\
\hline\hline
\end{tabular}
\end{table}


\section{Neutrino oscillation phase and probability}

A static, spherically symmetric configuration can be characterized by the following line element, which defines the geometry of spacetime:
\ie
\label{maininin}
\mathrm{d}s^{2} = f_{\Theta}(r)\mathrm{d}t^2-\dfrac{\mathrm{d}r^2}{f_{\Theta}(r)}-r^2\left(\mathrm{d}\theta^2+\sin^2\theta\mathrm{d}\varphi^2\right).
\fe
In a spacetime exhibiting spherical symmetry, as defined by the metric (\ref{maininin}), the dynamics of neutrinos occupying the \(k\)--th eigenstate are determined through the Lagrangian formulation given in \cite{neu18}:
\begin{align}
\mathcal{L}
& = \frac{1}{2}  m_{k} f_{\Theta}(r) \left(\frac{\mathrm{d}t}{\mathrm{d}\tau}\right)^2-\frac{1}{2}\frac{m_{k}}{f_{\Theta}(r)}\left(\frac{\mathrm{d}r}{\mathrm{d}\tau}\right)^2 \notag\\
& 
-\dfrac{1}{2}m_{k}r^2\left(\frac{\mathrm{d}\theta}{\mathrm{d}\tau}\right)^2  
 -\frac{1}{2}m_{k}r^2\sin^2\theta\left(\frac{\mathrm{d}\varphi}{\mathrm{d}\tau}\right)^2.
\end{align}

The canonical conjugate momentum associated with the coordinate $x^{\mu}$ is expressed as $p_{\mu} = \frac{\partial\mathcal{L}}{\partial\left(\frac{\mathrm{d}x}{\mathrm{d}\tau}\right)}$, where $\tau$ denotes the proper time and $m_k$ refers to the mass of the $k$--th eigenstate. By restricting the particle’s trajectory to the equatorial plane, $\theta = \frac{\pi}{2}$, the resulting nonvanishing momentum components are identified as follows \cite{neu60,Shi:2024flw}:
\begin{align}
p^{(k)t} &= m_{k}f_{\Theta}(r)\frac{\mathrm{d}t}{\mathrm{d}\tau} = E_{k}, \\
p^{(k)r} &= \frac{m_{k}}{f_{\Theta}(r)}\frac{\mathrm{d}r}{\mathrm{d}\tau}, \\
p^{(k)\varphi} &= m_{k}r^2\frac{\mathrm{d}\varphi}{\mathrm{d}\tau} = J_{k},
\end{align}
where the mass of the \(k\)--th eigenstate satisfies the mass--shell relation \cite{neu54,neu55}
\ie
m_{k}^2 = g^{(\Theta)}_{\mu\nu}p^{(k)\mu} p^{(k)\nu}.
\fe

The investigation of neutrino flavor transitions in curved spacetime has often relied on the plane wave approach, primarily when dealing with regions of weak gravitational influence \cite{neu53,neu54}. In the context of weak interactions, neutrinos manifest and are detected in their flavor states, rather than in mass eigenstates, as outlined in \cite{neu61,neu62,Shi:2024flw,neu63}
\ie
\ket{\nu_{\alpha}} = \sum_{i=1}^{3} \mathrm{U}_{\alpha i}^{*}\ket{\nu_{i}}.
\fe

The three neutrino flavors—electron, muon, and tau—are indexed by $\alpha = e, \mu, \tau$, while their corresponding mass eigenstates are denoted by $\ket{\nu_i}$. The transition between these two representations is defined through the unitary mixing matrix $\mathrm{U}$, which is $3 \times 3$ in dimension \cite{neu41}. Rather than treating neutrino propagation in terms of flavor alone, each mass eigenstate is described by a distinct wave function that evolves across spacetime. To streamline the notation, spacetime coordinates $\left(t_{\Tilde{S}}, \bm{x}_{\Tilde{S}}\right)$ and $\left(t_{\Tilde{D}}, \bm{x}_{\Tilde{D}}\right)$ are assigned to the emission point at the source ($\Tilde{S}$) and the detection point at the detector ($\Tilde{D}$), respectively. The evolution of the wave function along this trajectory is written as: 
\ie
\ket{\nu_{i}\left(t_{\Tilde{D}},\bm{x}_{\Tilde{D}}\right)} = e^{-\mathbbm{i}\Phi_{i}}\ket{\nu_{i}\left(t_{\Tilde{S}},\bm{x}_{\Tilde{S}}\right)},
\fe
such that the corresponding phase is expressed as
\ie
\Phi_{i}=\int_{\left(t_{\Tilde{S}},\bm{x}_{\Tilde{S}}\right)}^{\left(t_{\Tilde{D}},\bm{x}_{\Tilde{D}}\right)}g_{\mu\nu}^{(\Theta)}p^{(i)\mu}\mathrm{d}x^{\nu}.
\fe

The process of flavor oscillation is reconsidered here, focusing on the evolution of a neutrino as it travels from the source, where it is produced, to the detector, where it is measured. The probability of observing a transition from an initial flavor state $\nu_{\alpha}$ to a different flavor $\nu_{\beta}$ upon detection is given by the following probability:
\ie
\begin{split}
\mathrm{P}_{\alpha\beta}
& = |\left\langle \nu_{\beta}|\nu_{\alpha}\left(t_{\Tilde{D}}, \bm{x}_{\Tilde{D}}\right)\right\rangle|^2 \\
& = \sum_{i,j=1}^3 \mathrm{U}_{\beta i}\mathrm{U}_{\beta j}^{*} \mathrm{U}_{\alpha j} \mathrm{U}_{\alpha i}^{*}\,  e^{-\mathbbm{i}(\Phi_{i}-\Phi_{j})}.
\end{split}
\fe

The behavior of neutrinos restricted to the equatorial plane $(\theta = \frac{\pi}{2})$ is examined in the context of a gravitational background generated by a non--commutative black hole. Under these conditions, the corresponding phase takes the form:{
\begin{align}
\label{Pgefhi}
\Phi_{k} & = \int_{\left(t_{\Tilde{S}},\bm{x}_{\Tilde{S}}\right)}^{\left(t_{\Tilde{D}}, \bm{x}_{\Tilde{D}}\right)} g^{(\Theta)}_{\mu\nu} p^{(k)\mu}\mathrm{d}x^{\nu}\notag\\
& = \int_{\left(t_{\Tilde{S}},\bm{x}_{\Tilde{S}}\right)}^{\left(t_{\Tilde{D}}, \bm{x}_{\Tilde{D}}\right)}\left[E_{k}\mathrm{d}t - p^{(k)r}\mathrm{d}r-J_{k}\mathrm{d}\varphi\right] \notag\\
&
= \pm\frac{m_{k}^2}{2E_0}\int_{r_{\Tilde{S}}}^{r_{\Tilde{D}}}\left\{1-\frac{b^2}{r^2}\left[f_{\Theta}(r)\right]\right\}^{-\frac{1}{2}}\mathrm{d}r \notag\\
&\approx \pm \frac{m_{k}^2}{2E_0}\Biggl\{\sqrt{r_{\Tilde{D}}^2 - b^2}-\sqrt{r_{\Tilde{S}}^2 - b^2}\notag\\
&
+ \dfrac{M}{\sqrt{\pi}}\left[\dfrac{\sqrt{\pi}r_{\Tilde{D}}-4\sqrt{\Theta}}{\sqrt{r_{\Tilde{D}}^2 - b^2}}-\frac{\sqrt{\pi}r_{\Tilde{S}}-4\sqrt{\Theta}}{\sqrt{r_{\Tilde{S}}^2 - b^2}}\right] \notag\\
&
+ \dfrac{8M\sqrt{\Theta}}{b\sqrt{\pi}}\Biggr[\arctan\left(\dfrac{r_{\Tilde{D}}-\sqrt{r_{\Tilde{D}}^2-b^2}}{b}\right)\notag\\
&-\arctan\left(\dfrac{r_{\Tilde{S}}-\sqrt{r_{\Tilde{S}}^2-b^2}}{b}\right)\Biggl]\Biggl\}.
\end{align}
}

In this framework, the quantity  $E_0 = \sqrt{E_k^2 - m_k^2}$ characterizes the mean energy of relativistic neutrinos originating from the source, while the parameter $b$ refers to the impact parameter \cite{neu18}. As the neutrinos traverse the curved spacetime, their paths reach a point of closest approach at radius $r = r_0$. Within the regime of weak gravitational fields, this minimal distance $r_0$ is obtained by solving the orbital equation that dictates the neutrino’s trajectory
{\begin{align}
\label{r0}
r_0 \simeq \dfrac{b}{2}-\dfrac{M}{2}+\dfrac{1}{2}\sqrt{(M-b)^2+\dfrac{16M\sqrt{\Theta}}{\sqrt{\pi}}}.
\end{align}}

It is worth mentioning that the phase developed by a neutrino as it travels from the source, passes the point of minimum radial distance, and reaches the detector is derived by applying Eq. (\ref{Pgefhi}) together with the relation for $r_0$ specified in Eq. (\ref{r0}).
{\begin{align}
\label{pphhiii}
&\Phi_{k}\left(r_{\Tilde{S}}\to r_{0} \to r_{\Tilde{D}}\right)\notag\\
&\simeq \frac{{m}_{k}^2}{2E_0}
\Biggl\{
\sqrt{r_{\Tilde{S}}^2 - r_0^2}+\sqrt{r_{\Tilde{D}}^2 - r_0^2} \notag\\
& + M\left(\sqrt{\dfrac{r_{\Tilde{S}}-r_0}{r_{\Tilde{S}}+r_0}}+\sqrt{\dfrac{r_{\Tilde{D}}-r_0}{r_{\Tilde{D}}+r_0}}\right)\notag\\
&-\dfrac{4M\sqrt{\Theta}}{\sqrt{\pi}r_0}\Biggr[\pi-2\arctan\left(\dfrac{r_{\Tilde{S}}-\sqrt{r_{\Tilde{S}}^2-b^2}}{b}\right)\notag\\
&-2\arctan\left(\dfrac{r_{\Tilde{D}}-\sqrt{r_{\Tilde{D}}^2-b^2}}{b}\right)\Biggl]\Biggr\}.
\end{align}}

By performing a series expansion of Eq. (\ref{pphhiii}) up to terms of order $\frac{b^2}{r_{\Tilde{S},\Tilde{D}}^2}$, assuming the condition $b \ll r_{\Tilde{S},\Tilde{D}}$ holds, the resulting expression becomes:
{\begin{align}
\Phi_{k}
& \simeq \dfrac{{m}_{k}^2}{2E_0}
\Biggl\{\left(r_{\Tilde{S}} + r_{\Tilde{D}}\right) \Biggl[\left(1 - \dfrac{b^2}{2r_{\Tilde{S}} \, r_{\Tilde{D}}}\right)\notag\\
& +\dfrac{2M}{r_{\Tilde{S}} + r_{\Tilde{D}}}\left(1-\dfrac{2\sqrt{\pi\Theta}}{b}\right)\Biggr]\Biggr\}.
\end{align}}

As the non--commutative parameter increases, a clear modification arises: the phase accumulated during neutrino propagation diminishes with growing $\Theta$. The parameters employed in this analysis are $E_0 = 10\,\mathrm{MeV}, \quad r_{\Tilde{D}} = 10\,\mathrm{km}, \quad \text{and} \quad r_{\Tilde{S}} = 10^5 r_{\Tilde{D}}.$

As neutrinos propagate through curved spacetime, gravitational lensing influences their trajectories. To examine the flavor oscillation probability discussed earlier near the black hole, it becomes necessary to evaluate the phase difference accumulated along the distinct possible paths \cite{Shi:2024flw}.
\begin{align}
\Delta\Phi_{ij}^{pq}
&= \Phi_i^{p}-\Phi_j^{q}\notag\\
&= \left(\Delta m_{ij}^2 A_{pq}+\Delta b_{pq}^2 B_{ij}\right),
\end{align}
where
{\begin{align}
\Delta m_{ij}^2 & = m_i^2 - m_j^2,\\
\Delta b_{pq}^2 & = b_{p}^2-b_{q}^2,\\
A_{pq} & = \frac{r_{\Tilde{S}} + r_{\Tilde{D}}}{2 E_0} \Biggl\{1 -\frac{\sum b_{pq}^2}{4r_{\Tilde{S}} \, r_{\Tilde{D}}} \notag\\
&+\dfrac{2M}{r_{\Tilde{D}}+r_{\Tilde{S}}}\Biggl[1-\sqrt{\pi\Theta}\left(\dfrac{1}{b_p}+\dfrac{1}{b_q}\right)\Biggr]\Biggr\},\\
B_{ij} & = -\frac{\sum m_{ij}^2}{8E_0}\left[\frac{1}{r_{\Tilde{S}}} + \frac{1}{r_{\Tilde{D}}}-\dfrac{8M\sqrt{\pi\Theta}}{b_pb_q\left(b_p+b_q\right)}\right],\\
\sum b_{pq}^2 & = b_{p}^2 + b_{q}^2,\\
\sum m_{ij}^2 & = m_i^2 + m_j^2.
\end{align}}

To distinguish the phases associated with different neutrino trajectories, superscripts such as $\Phi_{i}^{p}$ are used, where each index $p$ identifies a distinct path characterized by its corresponding impact parameter $b_p$. The resulting phase difference that contributes to the neutrino transition probability in the presence of a non--commutative black hole depends on the individual neutrino masses $m_i$, the mass-squared differences $\Delta m_{ij}^2$, and the features of the gravitational background. When the non--commutative parameter $\Theta$ is set to zero, the expression for the phase difference recovers the standard result found in Ref. \cite{neu53}.

{The term $B_{ij}$ carries the dependence on the absolute neutrino mass scale through the combination $\sum m_{ij}^2 = m_i^2 + m_j^2$, whereas the influence of non--commutativity manifests as $\sqrt{\Theta}$--dependent corrections to both $A_{pq}$ and $B_{ij}$, which modify the accumulated phase and the amplitude of oscillation. At the order retained here, these $\Theta$ corrections are entirely absorbed into $A_{pq}$ and $B_{ij}$, so that no separate, irreducible path--mass coefficient survives; both $A_{pq}$ and $B_{ij}$ are symmetric under the interchange of their respective indices.}

{It should be emphasized that sensitivity to the absolute mass scale is not a generic feature of the oscillation probability. While individual phases $\Phi_k$ depend on $m_k^2$, the experimentally relevant phase difference $\Delta\Phi_{ij}^{pq}$ depends on the absolute scale only through the $\Delta b_{pq}^2\, B_{ij}$ contribution, which is proportional to $\sum m_{ij}^2$ and therefore requires genuinely distinct lensing paths, $b_p \neq b_q$. In the single--path limit $\Delta b_{pq}^2 \to 0$, the phase difference reduces to $\Delta m_{ij}^2\, A_{pq}$, recovering the standard result that observable probabilities are controlled solely by the mass--squared differences $\Delta m_{ij}^2$. The absolute--mass dependence reported below is thus a genuine but inherently multi--path effect, in line with the Schwarzschild analysis of Ref.~\cite{neu53}.}


\section{Neutrino gravitational lensing}

\begin{figure}
    \centering
    \includegraphics[scale=0.45]{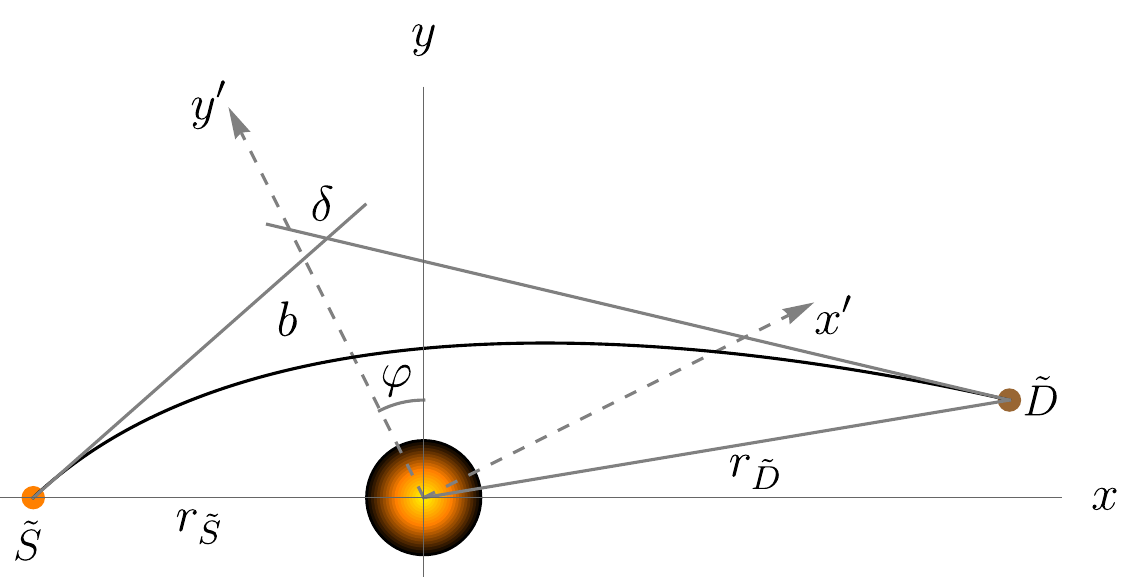}
    \caption{Schematic illustration of weak gravitational lensing affecting neutrino trajectories. In this diagram, $\Tilde{S}$ denotes the neutrino source, while $\Tilde{D}$ corresponds to the detection point.}
    \label{lensing}
\end{figure}

In the presence of a strong gravitational field generated by a massive object, neutrinos may follow nonradial trajectories, giving rise to gravitational lensing effects between the emission point and the detector \cite{neu54}. This lensing allows neutrinos traveling along multiple distinct paths to arrive at the same detection point $D$ (illustrated in Fig. \ref{lensing}). Consequently, the flavor eigenstate of the neutrino must be redefined to incorporate contributions from all relevant paths \cite{Shi:2024flw,neu56,neu62,neu63,neu64,neu65}:
\ie
|\nu_{\alpha}(t_{\Tilde{D}},x_{\Tilde{D}})\rangle = N\sum_{i}\mathrm{U}_{\alpha i}^{\ast}
\sum_{p} e^{- \mathbbm{i} \Phi_{i}^{p}}|\nu_{i}(t_{\Tilde{S}}, x_{\Tilde{S}})\rangle,
\fe
where $p$ labels the distinct paths taken by the neutrinos. Given that virtually all trajectories intersect at the detector location, the total probability for observing a flavor transition $\nu_{\alpha} \rightarrow \nu_{\beta}$ upon detection is expressed as \cite{Shi:2024flw,neu56,neu62,neu63,neu64,neu65}:
\begin{align}
\label{nasndkas}
\mathcal{P}_{\alpha\beta}^{\mathrm{lens}} & = |\langle \nu_{\beta}|\nu_{\alpha}(t_{\Tilde{D}}, x_{\Tilde{D}})\rangle|^{2}\notag\\
& =|N|^{2}\sum_{i, j}\mathrm{U}_{\beta i}\mathrm{U}_{\beta j}^{\ast}\mathrm{U}_{\alpha j}\mathrm{U}_{\alpha j}^{\ast}\sum_{p, q}e^{\Delta\Phi_{ij}^{pq}},
\end{align}
which leads to the following expression for the normalization constant:
\ie
|N|^{2} = \Biggl[\sum_{i}|\mathrm{U}_{\alpha i}|^{2}\sum_{p,q}e^{(-\mathbbm{i}\Delta\Phi_{ij}^{pq})}\Biggl]^{-1}.
\fe

Taking into account the phase difference $\Delta\Phi_{ij}^{pq}$ introduced in earlier expressions, the probability of neutrino oscillation under the influence of gravitational lensing is shaped by multiple elements—including the individual masses of the neutrinos, the mass-squared differences, and the specific properties of the black hole spacetime, as outlined in Eq. (\ref{nasndkas}). This pattern resembles the behavior observed around spherically symmetric backgrounds such as the Schwarzschild solution \cite{neu53}.

The analysis now turns to the impact of gravitational lensing on neutrino oscillation probabilities, with particular attention to the role played by the non--commutative parameter $\Theta$. In scenarios where non--commutativity acts as a lensing mechanism for two--flavor neutrinos, the transition probability $\nu_{\alpha} \to \nu_{\beta}$ at the detector is studied. This probability is derived within the weak--field approximation, considering the spatial configuration defined by the positions of the source, the lens, and the detector \cite{neu53,neu54,neu55,Shi:2024flw,neu65}
\begin{align}
\label{asdPdadbd2}
\mathcal{P}_{\alpha\beta}^{\mathrm{lens}}
&= \left|N\right|^2\biggl\{2\sum_i\left|U_{\beta i}\right|^2\left|U_{\alpha i}\right|^2\left[1+\cos\left(\Delta b_{12}^2B_{ii}\right)\right]\notag\\
&+\sum_{i\neq j}U_{\beta i}U_{\beta j}^*U_{\alpha j}U_{\alpha i}^*\left(e^{-\mathbbm{i}\Delta m_{ij}^2 A_{11}}+e^{-\mathbbm{i}\Delta m_{ij}^2 A_{22}}\right)\notag\\
&+\sum_{i\neq j}U_{\beta i}U_{\beta j}^*U_{\alpha j}U_{\alpha i}^*e^{-\mathbbm{i}\Delta b_{12}^2B_{ij}}e^{-\mathbbm{i}\Delta m_{ij}^2A_{12}}\notag\\
&+\sum_{i\neq j}U_{\beta i}U_{\beta j}^*U_{\alpha j}U_{\alpha i}^*e^{\mathbbm{i}\Delta b_{21}^2B_{ij}}e^{-\mathbbm{i}\Delta m_{ij}^2A_{21}}\biggr\}.
\end{align}

The structure of the probability formula in Eq. (\ref{asdPdadbd2}) includes several contributions enclosed within curly brackets, each corresponding to specific configurations of mass and path indices. The first term arises when $i = j$, representing diagonal elements linked to the same mass eigenstate. The second term is associated with interference between different mass states along the same trajectory, i.e., $i \neq j$ and $p = q$. The remaining two terms deal with cases where both the mass eigenstates and the propagation paths differ ($i \neq j$, $p \neq q$), with separate treatment for $p < q$ and $p > q$.

{Although the formalism developed above is fully three--flavor, the numerical analysis that follows is carried out in an effective two--flavor scheme. This is a standard and well--controlled approximation: for the energies and baselines considered here a single mass--squared difference and a single mixing angle dominate the interference pattern, so that the essential physics---and in particular the $\Theta$--induced coherent residual that is the focus of this work---is already captured at the two--flavor level. The reduction does not alter our qualitative conclusions. As an explicit check, the full three--flavor residuals for all channels ($\nu_e\to\nu_\mu$, $\nu_e\to\nu_\tau$, $\nu_\mu\to\nu_\tau$) and for both mass orderings are reported in Appendix~\ref{app:3flavor}, where the same oscillatory $\Theta$--signature is found to persist.}

For the case of two neutrino flavors, the mixing matrix reduces to a $2 \times 2$ unitary matrix governed solely by the mixing angle $\alpha$ \cite{neu43}
\begin{align}
\label{U}
\mathrm{U}\equiv\left(\begin{matrix}
\cos\alpha&\sin\alpha\\
-\sin\alpha&\cos\alpha
\end{matrix}\right).
\end{align}

Replacing the mixing matrix from Eq. (\ref{U}) into the general expression for transition probability in Eq. (\ref{asdPdadbd2}) yields the specific form of the oscillation probability for the process $\nu_e \to \nu_\mu$ as follows
\begin{align}
\label{proballl}
\mathcal{P}_{\alpha\beta}^{\mathrm{lens}}
&=\left|N\right|^2\sin^{2}2\alpha\notag\\
&\times\biggl[\sin^2\left(\dfrac{1}{2}\Delta m_{12}^2A_{11}\right)+\sin^2\left(\dfrac{1}{2}\Delta m_{12}^2A_{22}\right)\notag\\
&+\dfrac{1}{2}\cos\left(\Delta b_{12}^2B_{11}\right)+\dfrac{1}{2}\cos\left(\Delta b_{12}^2B_{22}\right)\notag\\
&\quad-\cos\Delta b_{12}^2B_{12}\cos\Delta m_{12}^2A_{12}\biggr].
\end{align}

Considering the leptonic mixing matrix given in Eq. (\ref{U}) along with the phase differences accumulated through the various neutrino trajectories, the normalization constant takes the following form:
\begin{align}
\left|N\right|^2&=\biggl[2+2\cos^2\alpha\cos\left(\Delta b_{12}^2B_{11}\right)\notag\\
&+2\sin^2\alpha\cos\left(\Delta b_{12}^2B_{22}\right)\biggr]^{-1}.
\end{align}


\section{Numerical analysis}

To clarify the behavior of neutrino oscillations in the black hole spacetime considered here, it is essential to analyze the lensing probabilities presented in Eq. (\ref{proballl}). In the adopted $(x, y)$ coordinate system, the lens is positioned at the origin, while the neutrino source and detector are separated from the lens by the physical distances $r_{\Tilde{S}}$ and $r_{\Tilde{D}}$, respectively. One can also define a rotated coordinate system $(x', y')$, obtained from $(x, y)$ via a rotation by an angle $\varphi$, such that  
\ie 
x' = x\cos\varphi + y\sin\varphi, \quad y' = -x\sin\varphi + y\cos\varphi 
\nonumber
\fe
\cite{neu53,Shi:2024flw}. When $\varphi = 0$, all three components of the setup—the source, lens, and detector—lie along a single straight line in the plane.

Following Refs. \cite{neu53,Shi:2024flw}, the impact parameter $b$ and the deflection angle $\delta$, which quantifies the deviation of the neutrino from its original path due to gravitational lensing, are related by:
{\begin{align}
\label{delta}
\delta \sim \dfrac{y_{\Tilde{D}}'- b}{x_{\Tilde{D}}'}=-\dfrac{4M}{b}+\dfrac{6M\sqrt{\pi\Theta}}{b^2}
\end{align}}
with the detector positioned at $(x_{\Tilde{D}}', y_{\Tilde{D}}')$ in the rotated coordinate frame. Using the identity $\sin\varphi = \frac{b}{r_S}$, the expression in Eq. (\ref{delta}) can be reformulated as:
{\begin{align}
\label{solve_b}
&\left(4Mbx_{\Tilde{D}}-6M\sqrt{\pi\Theta}x_{\Tilde{D}}+b^2y_{\Tilde{D}}\right)\sqrt{1-\dfrac{b^2}{r_{\Tilde{S}}^2}}\notag\\
&=b^3\left(\dfrac{x_{\Tilde{D}}}{r_{\Tilde{S}}}+1\right)-\left(4Mb-6M\sqrt{\pi\Theta}\right)\dfrac{by_{\Tilde{D}}}{r_{\Tilde{S}}}.
\end{align}}

We now proceed to evaluate the lensing probability of neutrino oscillation in the presence of a black hole, aiming to highlight the influence of the non--commutative parameter $\Theta$. A meaningful comparison can be made with the results obtained for the Schwarzschild black hole in Ref. \cite{swami2020signature}. The relevant impact parameters—such as $r_{\Tilde{S}}$, $r_{+}$, and the coordinates of the lensing point $(x_{\Tilde{D}}, y_{\Tilde{D}})$—can be determined using Eq. (\ref{delta}).

\begin{figure*}
\centering
\includegraphics[width=13cm]{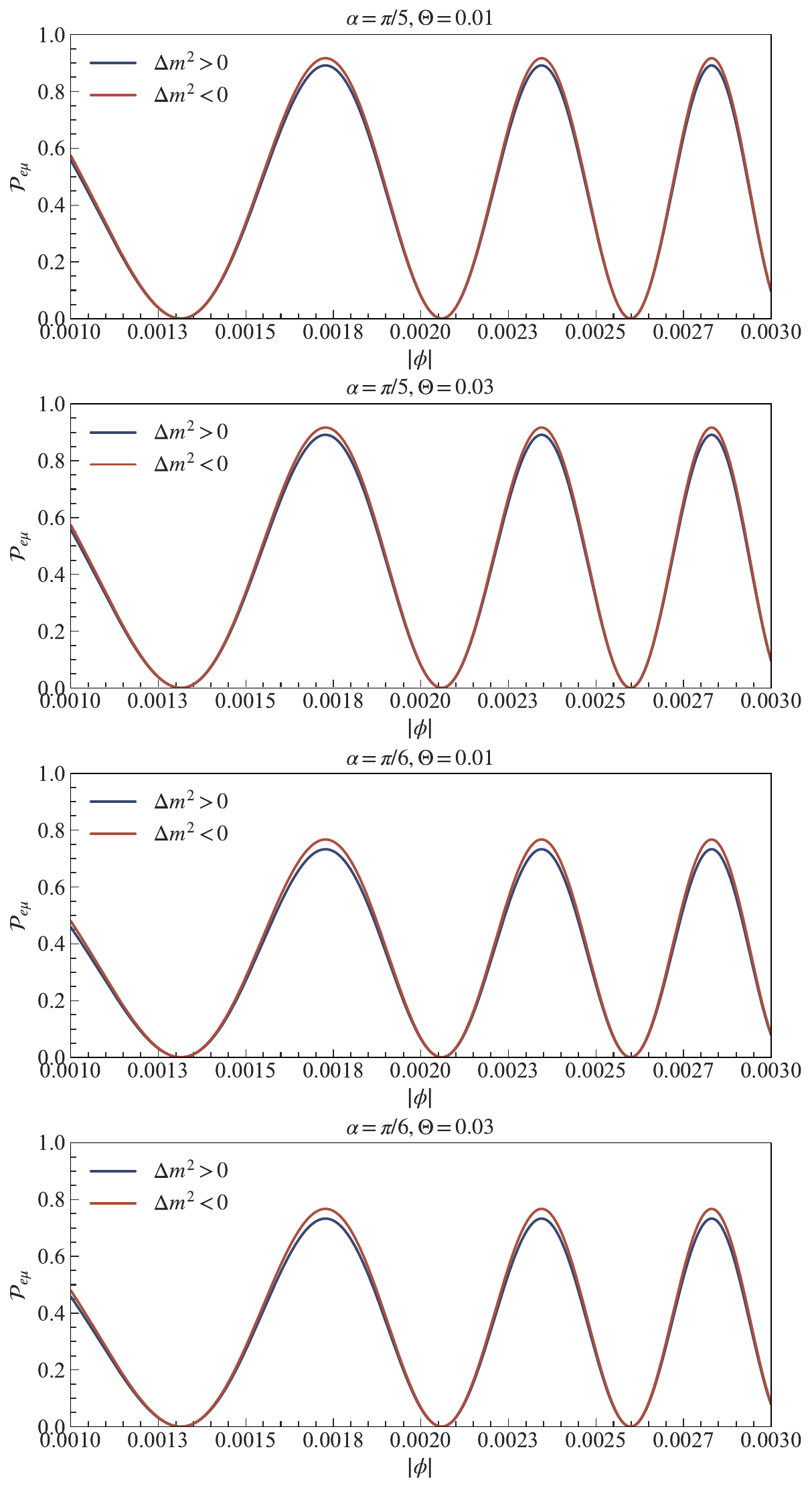}
\caption{\label{fig:prob1} {For $\Theta = 0.01, 0.03$, oscillation probability $\mathcal{P}_{e\mu}$ vs. azimuthal angle $|\phi|$ for normal and inverted mass orderings. $\alpha = \pi/5, \pi/6$ are the mixing angles. At this scale, deviations from the Schwarzschild backdrop are visually insignificant.}}
\end{figure*}

The overall behavior of neutrino flavor transitions $\nu_e \to \nu_\mu$ as a function of the azimuthal angle $\varphi \in [0, 0.003]$ is shown in Figure \ref{fig:prob1}. In this preliminary study, the lightest neutrino is taken to be massless. The findings for non-commutative parameters $\Theta = 0.01$ and $\Theta = 0.03$ under two mixing angle situations, $\alpha = \frac{\pi}{5}$ and $\alpha = \frac{\pi}{6}$, are compared in the figure. Positive values of $\Delta m^2$ are represented by the blue curves (normal ordering), whilst negative values are represented by the red curves (inverted ordering).

The inverted mass ordering consistently produces greater transition probabilities than the usual ordering across all panels, as seen in Fig. \ref{fig:prob1}. This finding demonstrates that the sign of $\Delta m^2$ continues to have a significant impact on the gravitational lensing effect on neutrino oscillation. It is crucial to remember that the deviations caused by the non-commutative parameter $\Theta$ are visually identical to the Schwarzschild background at this macroscopic size. In order to separate the non-commutative corrections, this seemingly insignificant difference requires a more thorough high-precision residual analysis, which will be explained in the following figures.

\begin{figure*}
\centering
\includegraphics[width=13cm]{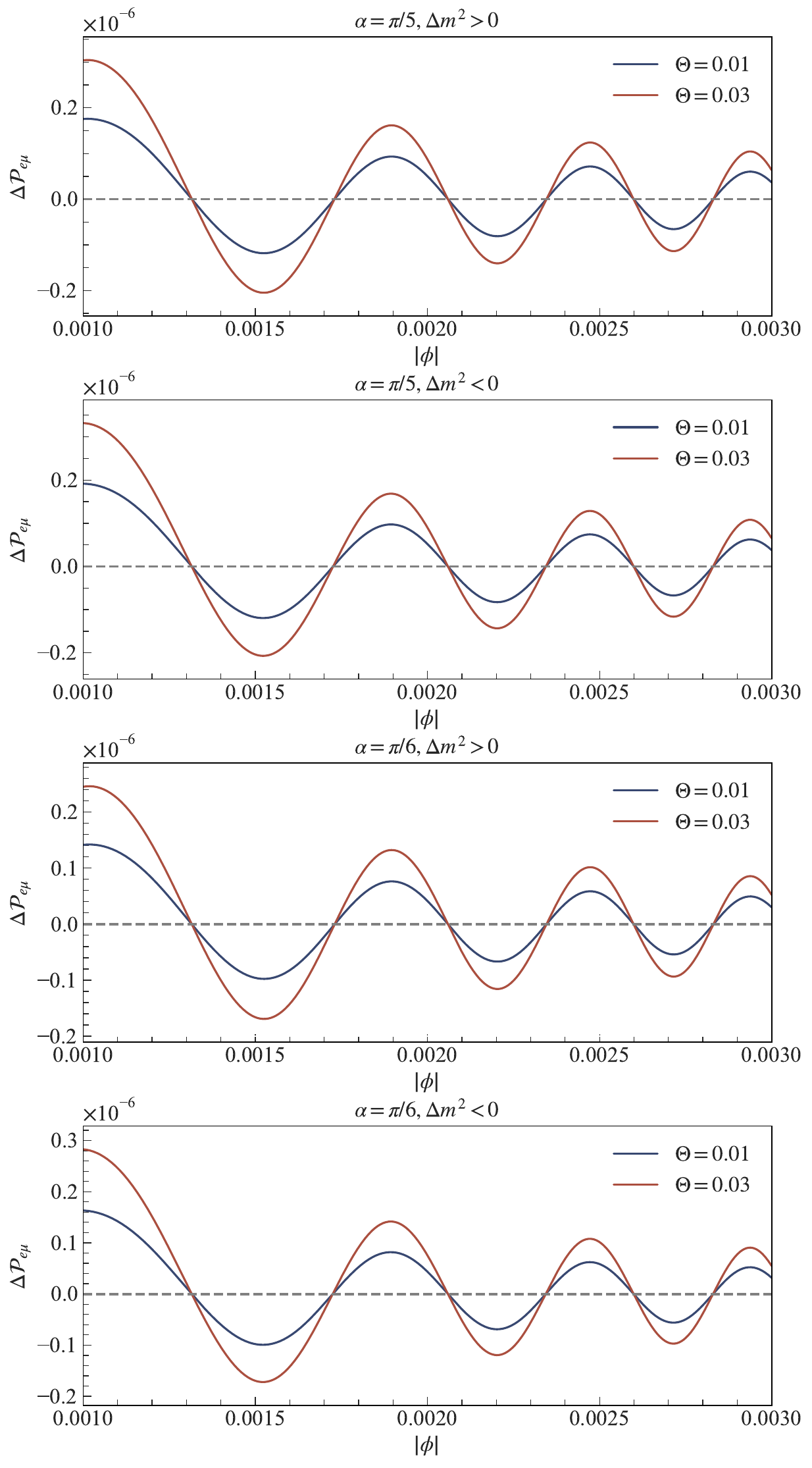}
\caption{\label{fig:prob2} {For $\Theta = 0.01$ (blue) and $0.03$ (red), the residual probability $\Delta \mathcal{P}_{e\mu} = \mathcal{P}_{\Theta} - \mathcal{P}_{\mathrm{Sch}}$. A coherent phase shift generated by $\Theta$ is shown by the unique oscillating patterns of order {$10^{-7}$}, proving that ordinary Schwarzschild geometry cannot replicate non-commutative effects.}}
\end{figure*}

We conducted a high-precision numerical analysis to separate the unique contribution of the non-commutative parameter $\Theta$ from the Schwarzschild background. The residual probability, $\Delta \mathcal{P}_{e\mu} = \mathcal{P}_{\Theta} - \mathcal{P}_{\mathrm{Sch}}$, is shown in Figure \ref{fig:prob2}. The residual plots show a characteristic, non-zero oscillating pattern with an amplitude on the order of $10^{-6}$, in contrast to the macroscopic picture in Fig. \ref{fig:prob1}.

In order to break the parameter degeneracy, this structure is essential. The {residual} would be zero or a constant offset if the non-commutative effect were only imitating a Schwarzschild black hole with a different mass. Rather, the distinct rhythmic pattern suggests that $\Theta$ causes a distinct, coherent phase shift in the neutrino wavefunction. Additionally, we find that the amplitude of the residual signal grows monotonically with $\Theta$ when comparing the blue curves ($\Theta=0.01$) and red curves ($\Theta=0.03$), verifying the perturbation's physical consistency.

\begin{figure*}
\centering
\includegraphics[width=13cm]{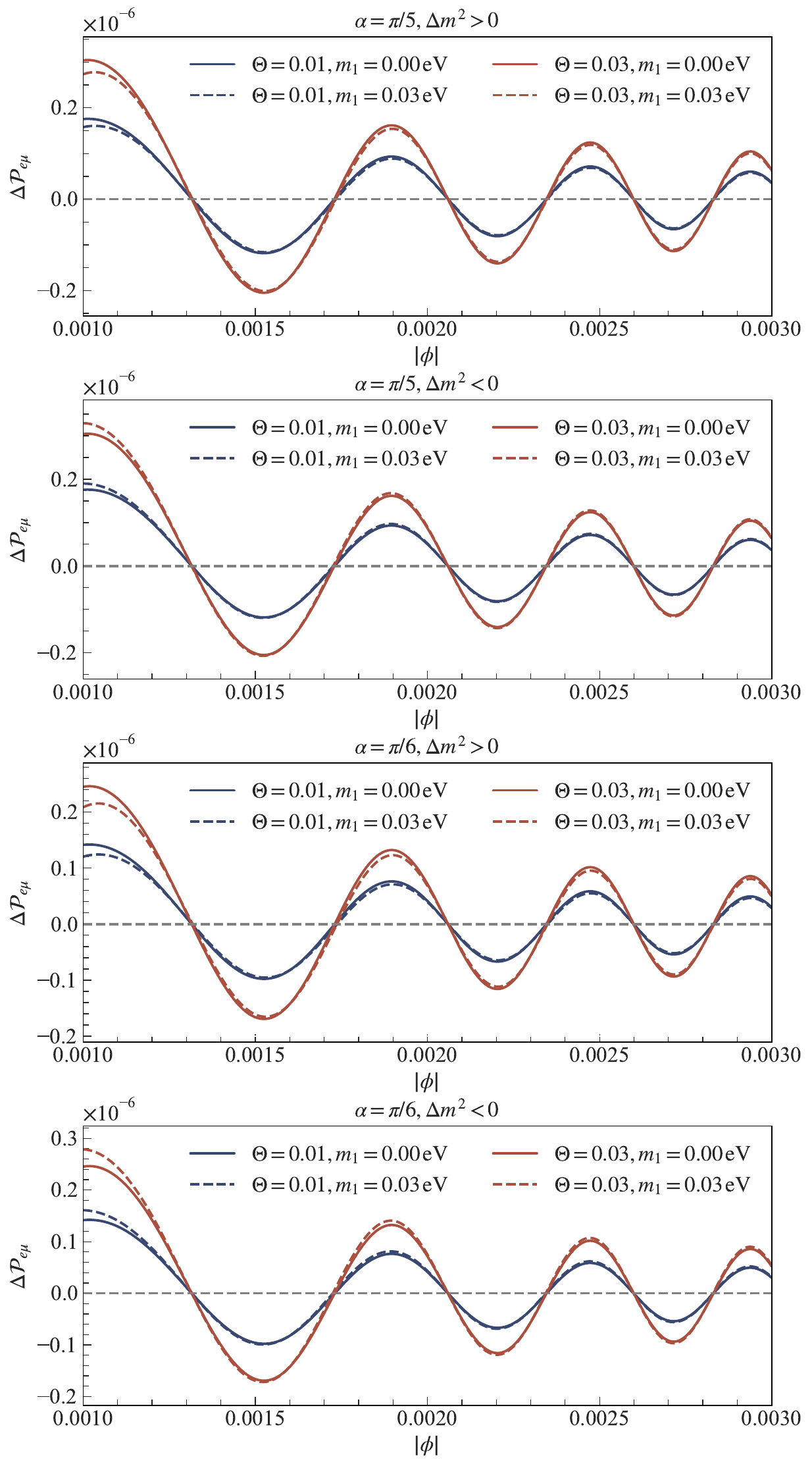}
\caption{\label{fig:prob3} {Dependence of the residual signal on the absolute neutrino mass scale. The curves for $m_1 = 0.00$ (solid) and $m_1 = 0.03\,\mathrm{eV}$ (dashed) display a visible but subdominant offset that grows toward larger $\varphi$. This confirms that the non-commutative parameter $\Theta$, rather than the absolute mass scale, is the primary driver of the observed phase shift, the latter providing only a subleading modulation.}}
\end{figure*}

Lastly, we {examine} how this residual signal {depends} on the absolute neutrino mass scale. The residual probabilities under both normal and inverted mass orderings are compared for {two representative values of the} lightest neutrino mass scenarios in Figure \ref{fig:prob3}: $m_1 = 0.00\,\mathrm{eV}$ (solid lines), {and $m_1 = 0.03\,\mathrm{eV}$ (dashed lines)}.

{For a fixed value of $\Theta$, the residual curves corresponding to different values of $m_1$ exhibit a visible but clearly subdominant dependence on the absolute mass scale: the solid ($m_1=0$) and dashed ($m_1=0.03\,\mathrm{eV}$) curves track one another closely, developing only a small offset that grows toward larger $\varphi$. This is precisely what is expected from the structure of the phase difference, in which the absolute mass scale enters exclusively through the $\Delta b_{12}^2\, B_{ij}$ term: its imprint on the residual is genuine but remains secondary to the $\sqrt{\Theta}$--driven oscillatory pattern. The non-commutative geometry parameter $\Theta$ therefore remains the dominant driver of the coherent phase shift shown in Fig. \ref{fig:prob2}, with the absolute neutrino mass providing only a sub-leading modulation.}

{It is important to place the magnitude of this signal in observational context. The residual oscillation probability, of order $10^{-7}$, lies far below the sensitivity of current and foreseeable neutrino detectors, whose flavor--composition measurements reach at best the percent level---as, for instance, in the flavor--ratio determinations at high--energy neutrino observatories. Astrophysical uncertainties in the source energy spectrum, in the source--lens--detector alignment, and in the absolute distances would further dilute any such imprint. The residual analyzed here should therefore be understood as an \emph{in--principle}, geometric discriminator---a coherent signature that, by construction, cannot be reproduced by a Schwarzschild black hole with a redefined mass---rather than as a quantity directly measurable with present or near--future technology. Its value lies in establishing that non--commutativity leaves a qualitatively distinct, non--degenerate fingerprint on gravitationally lensed flavor oscillations.}

It is clearly observed that for a fixed value of $\Theta$, the transition probability curves vary depending on the value of the lightest neutrino mass—indicating that the oscillation behavior is sensitive not only to mass differences but also to the absolute mass scale. This dependence of the transition probability on the azimuthal angle $\varphi$ is evident across all panels and is influenced by the choice of mixing angle, either $\alpha = \frac{\pi}{5}$ or $\alpha = \frac{\pi}{6}$, as well as by the sign of $\Delta m^2$.

Similar to the case of the Schwarzschild black hole explored in Ref. \cite{swami2020signature}, the oscillation pattern varies with the individual neutrino mass. Moreover, significant differences in transition probabilities are found for certain azimuthal angles $\varphi$ when comparing results across different values of the non--commutative parameter $\Theta$, further emphasizing its impact on neutrino lensing and oscillation.

\section{Conclusion}

{

In this work, neutrino dynamics were investigated in a spherically symmetric non--commutative spacetime generated by a Lorentzian smeared matter distribution. The geometry was treated explicitly through the $\Theta$--dependent metric function, without introducing an effective mass redefinition, which allowed the direct impact of non--commutativity on physical observables to be identified.

The neutrino--antineutrino annihilation process was first analyzed. It was found that the relativistic energy deposition rate decreased monotonically as $\Theta$ increased. This suppression arose from the modified radial dependence of the metric functions entering the redshift relation and the angular integration factor. The Newtonian limit was recovered at large radii, while near the compact source the deviation from the Schwarzschild case became more pronounced. These results indicated that non--commutativity reduced the ``efficiency'' of neutrino--driven energy release mechanisms in strong gravitational fields.

The phase acquired by neutrino mass eigenstates along geodesics was then derived. In the weak--field regime, analytic expressions were obtained showing that $\Theta$ introduced corrections proportional to $\sqrt{\Theta}$, modifying both the minimal distance of approach and the accumulated phase. As $\Theta$ increased, the total phase decreased relative to the Schwarzschild background, leading to explicit corrections in the oscillation probability.

The gravitational lensing analysis further clarified this behavior. Although the macroscopic transition probabilities closely followed those of the Schwarzschild geometry, a high-precision residual analysis revealed a nonvanishing oscillatory correction of order {$10^{-7}$} in the transition probability. This residual pattern could not be reproduced by a simple redefinition of the mass parameter and originated from the $\Theta$--dependent structure of the phase difference between distinct trajectories. {Furthermore, variations of the absolute neutrino mass scale left the residual signal largely unchanged, modifying it only at a subdominant level, which demonstrated that the effect was primarily geometric rather than kinematical in origin. Owing to its small amplitude, this residual is best regarded as an in--principle, non--degenerate signature of non--commutativity rather than a presently measurable quantity.} Furthermore, variations of the absolute neutrino mass scale left the residual signal essentially unchanged, demonstrating that the effect was geometric rather than kinematical in origin.


}


\section*{Acknowledgments}
\hspace{0.5cm} A. A. Araújo Filho is supported by Conselho Nacional de Desenvolvimento Cient\'{\i}fico e Tecnol\'{o}gico (CNPq) and Fundação de Apoio à Pesquisa do Estado da Paraíba (FAPESQ), project numbers 150223/2025-0 and 1951/2025. N. H. would like to acknowledge networking support of the COST Action CA 22113 - Fundamental challenges in theoretical physics (Theory and Challenges), CA 21106 - COSMIC WISPers in the Dark Universe: Theory, astrophysics and experiments (CosmicWISPers), CA 21136 - Addressing observational tensions in cosmology with systematics and fundamental physics (CosmoVerse), and CA 23130 - Bridging high and low energies in search of quantum gravity (BridgeQG). I. P. L. was partially supported by the National Council for Scientific and Technological Development - CNPq, grant 312547/2023-4 and to acknowledge networking support by the COST Action BridgeQG (CA23130), the COST Action RQI (CA23115) and the COST Action FuSe (CA24101) supported by COST (European Cooperation in Science and Technology).

\appendix

\section{Three--flavor residual oscillations}
\label{app:3flavor}

\begin{figure*}
\centering
\includegraphics[width=13cm]{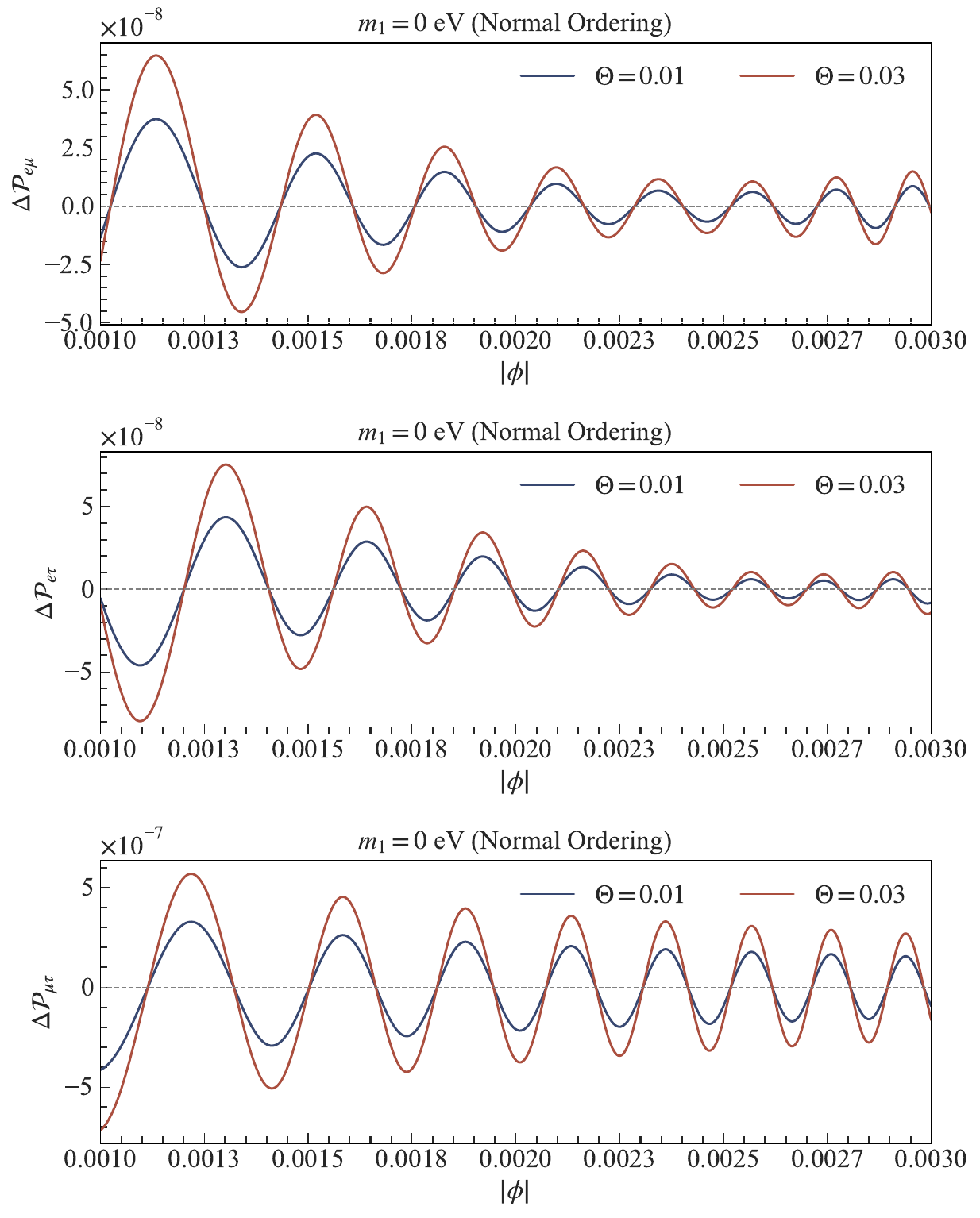}
\caption{\label{fig:3f_no} {Three--flavor residual probabilities $\Delta\mathcal{P}_{\alpha\beta} = \mathcal{P}_{\alpha\beta}^{\Theta} - \mathcal{P}_{\alpha\beta}^{\mathrm{Sch}}$ for the normal ordering ($m_1 = 0$), in the channels $\nu_e\!\to\!\nu_\mu$, $\nu_e\!\to\!\nu_\tau$ and $\nu_\mu\!\to\!\nu_\tau$ (top to bottom), with $\Theta = 0.01$ (blue) and $\Theta = 0.03$ (red).}}
\end{figure*}

\begin{figure*}
\centering
\includegraphics[width=13cm]{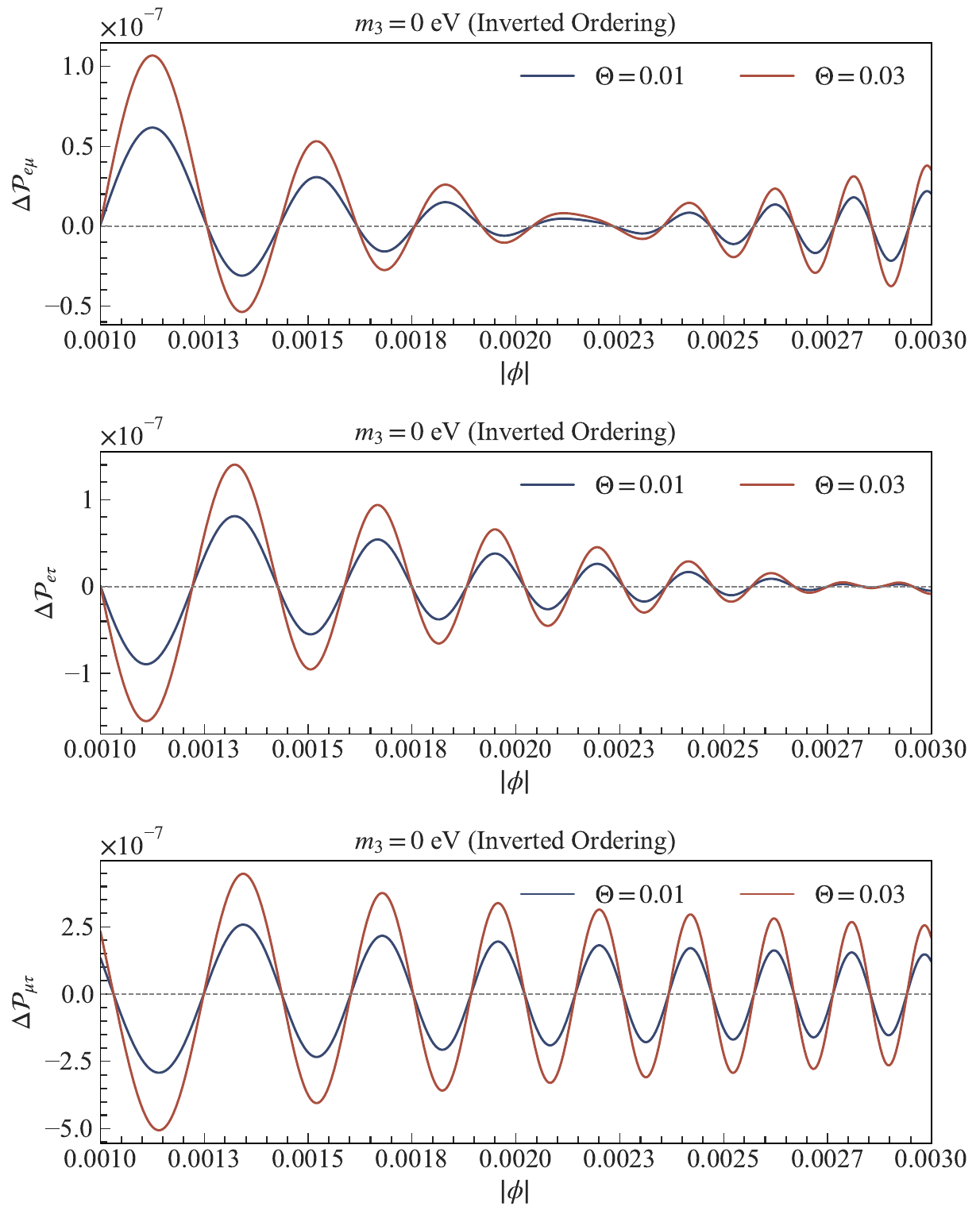}
\caption{\label{fig:3f_io} {Same as Fig.~\ref{fig:3f_no}, but for the inverted ordering ($m_3 = 0$).}}
\end{figure*}

{To verify that the effective two--flavor treatment adopted in the main text does not affect our conclusions, we repeat the gravitational--lensing analysis within the full three--flavor framework, using the standard PMNS matrix with central values of the oscillation parameters. As in the two--flavor case, the macroscopic three--flavor transition probabilities are visually indistinguishable from the Schwarzschild result; we therefore present directly the residuals $\Delta\mathcal{P}_{\alpha\beta} = \mathcal{P}_{\alpha\beta}^{\Theta} - \mathcal{P}_{\alpha\beta}^{\mathrm{Sch}}$. Figures~\ref{fig:3f_no} and \ref{fig:3f_io} show these residuals for all three channels and for the normal and inverted mass orderings, respectively. In every channel the residual retains the same coherent, non--vanishing oscillatory structure found in the two--flavor analysis, with an amplitude that grows with $\Theta$. This confirms that the non--commutative signature is a robust feature of the full three--flavor dynamics and is not an artifact of the two--flavor reduction.}

\bibliographystyle{ieeetr}
\bibliography{main}

\end{document}